\documentclass[reqno, 12pt]{amsart}

\numberwithin{equation}{section}

\usepackage{amssymb}
\usepackage{mathrsfs}
\usepackage{dsfont}
\usepackage{stmaryrd, MnSymbol}
\usepackage{enumerate, xspace}
\usepackage[bottom,first]{draftcopy}
\usepackage{bbm}
\usepackage{changebar}
\usepackage{hyperref}
\usepackage{pdfsync}
\usepackage{color}

\newtheorem{thmm}{Theorem}

\newtheorem{thm}{Theorem}[section]
\newtheorem{lem}[thm]{Lemma}

\newtheorem{prop}[thm]{Proposition}

\newcommand\cC{{\mathcal C}}

\newcommand\cO{{\mathcal O}}

\newcommand\bE{{\mathbb E}}

\newcommand\bP{{\mathbb P}}
\newcommand\bR{{\mathbb R}}

\newcommand\bZ{{\mathbb Z}}

\newcommand\ve{\varepsilon}

\newcommand{\mcE}{{\mc E\!\!\!\!\mc E\!\!\!\!\mc E}}

\newcommand{\mc}[1]{{\mathcal #1}}
\newcommand{\mf}[1]{{\mathfrak #1}}

\newcommand{\bb}[1]{{\mathbb #1}}

\newcommand{\triple}{|\!|\!|}

\newcommand{\tgamma}{\tilde{\gamma}}

\begin{document}

\title[Energy Model]{Diffusive Scaling in Energy Ginzburg-Landau Dynamics}
\author{Carlangelo Liverani}
\address{Carlangelo Liverani\\
Dipartimento di Matematica\\
II Universit\`{a} di Roma (Tor Vergata)\\
Via della Ricerca Scientifica, 00133 Roma, Italy.}
\email{{\tt liverani@mat.uniroma2.it}}
\author{Stefano Olla}
\address{Stefano Olla\\
CEREMADE, UMR CNRS 7534\\
Universit\'e Paris-Dauphine\\
75775 Paris-Cedex 16, France \\
}
\email{{\tt olla@ceremade.dauphine.fr}}
\author{Makiko Sasada}
\address{Makiko Sasada\\
University of Tokyo\\
Tokyo, Japan.}
\email{{\tt sasada@ms.u-tokyo.ac.jp}}

\date{\today. Preliminary Draft}
\begin{abstract}
Ginzburg-Landau energy models arise as autonomous stochastic dynamics
for the energies in coupled systems after a weak coupling limit
(cf. \cite{dolgoliv,lojams}). We prove here that, under certain
conditions, the energy fluctuations of these stochastic dynamics
 are driven by the heat equation, under a diffusive space time
 scaling.   
\end{abstract}
\thanks{This paper has been partially supported by the
  European Advanced Grant {\em Macroscopic Laws and Dynamical Systems}
  (MALADY) (ERC AdG 246953)} 
\maketitle


\section{Introduction}
\label{sec:introduction}

Heat equation describes the macroscopic evolution of energy, after a
coarse-graining limit, in system that have finite thermal
conductivity. In a pinned chain of anharmonic oscillators or other
coupled system conserving total energy, we expect that after a
diffusive rescaling of space and time the space distribution of the energy
converges to the solution of the corresponding heat equation. 
This is a very difficult open problem, even at the level of
equilibrium fluctuations, that should converge to the solution of the
corresponding linearized heat equation. 

In recent years some mathematical progress have been obtained for
weakly coupled very time mixing systems, in particular in
\cite{dolgoliv} for deterministic dynamics in negative curvature
manifolds, and in \cite{lojams} for an-harmonic oscillators with
 stochastic perturbations that conserve kinetic energy of each oscillator. 
In these weak coupling limit, an autonomous stochastic dynamics of the
energies of each system arise. These energies dynamics satisfy a
system of stochastic differential equations conservative of the total
energy and where the instantaneous energy exchange currents are
related to the equilibrium fluctuation variance of the corresponding
currents in the microscopic dynamics.   
These stochastic differential equations define a Markov process on the
energies configurations, and are formally similar to the
(non-gradient) Ginzburg-Landau dynamics considered by Varadhan in
\cite{Va}. The main differences with respect to the dynamics in
\cite{Va} are that the dynamics is here confined to positive values of
the energies, and that the families of stationary (reversible)
probability distributions correspond to potential growing linearly for
large values of the energy (since they are derived form the energies
marginals of the canonical Gibbs measures of the microscopic
dynamics). 
  
We obtain here the linearized heat equation for the behaviour of the
energy fluctuations in equilibrium under a diffusive space-time
scaling, for the energy Ginzburg-Landau dynamics. We use the
non-gradient approach of Varadhan used in \cite{Va}, properly
adapted. Consequently this result is conditioned to the existence of a
lower bound for the spectral gap of the generator of the corresponding
finite dynamics. This gap bound is proven in \cite{os}  for the Ginzburg-Landau
dynamics arising from the anharmonic oscillators with stochastic
perturbations considered in \cite{lojams}. For the GL arising in the
deterministic case considered by \cite{dolgoliv} it remains an open
problem, and in this moment we are not sure about the validity of such
bound. 

The present result imply a proof of the validity of the heat equation
in two step: first the weak-coupling limit (\cite{dolgoliv,lojams}),
then the hydrodynamic diffusive space-time limit (at least in the
linearized sense). A straight limit form the microscopic dynamics,
without passing through the weak-coupling, has been performed in
\cite{os} in a special situation with a stochastic perturbation of the
hamiltonian dynamics that involves directly also the positions. 

The difficulties we encounter already in the two step are certainly
instructive about the more difficult problem of the direct limit from
the microscopic dynamics.

\section{The dynamics}
\label{sec:dynamics}


We consider the dynamics for the infinite system, and, in order to
keep notation as simple as possible, in one dimension. 
The configuration space is given by:
$\mcE = \{\mc E_x , x\in \bb Z\} \in (\bb R_+)^{\bb Z}$.

The dynamics is defined by the solution of the stochastic differential
equations: 
\begin{equation}
  \label{eq:sde1}
  \begin{split}
    d \mc E_x(t) &= dJ_{x-1,x}(t) - dJ_{x+1,x}(t)\\
    dJ_{x+1,x}(t) &= \alpha(\mc E_x(t), \mc E_{x+1}(t)) dt + \sqrt 2
    \gamma(\mc E_x(t), \mc E_{x+1}(t)) dB_{x,x+1}
  \end{split}
\end{equation}
The coefficients are related by the equations 
\begin{equation}
  \label{eq:revcond}
  \alpha(\mc E_0, \mc E_{1}) = e^{\mc U(\mcE)} 
  \left(\partial_{\mc E_1} - \partial_{\mc E_0}\right) \left( e^{-\mc
      U(\mcE)} \gamma^2(\mc E_0, \mc E_{1}) \right) 
\end{equation}
with $\mc U(\mcE) = \sum_x U(\mc E_x)$, and we are interested in $U(a)
\sim \log a$ for $a\to 0$ and $a\to\infty$. We will specify further conditions
on $\gamma^2$. 
Sometimes we will use the notation $\alpha_{x,x+1} = \alpha(\mc E_x,
\mc E_{x+1})$, and similarly for $\gamma_{x,x+1}$.

The corresponding generator can be formally written as 
\begin{equation}
  \label{eq:gen}
  \begin{split}
    L &= \sum_x L_{x,x+1} \\
    L_{x,x+1} &= e^{\mc U(\mcE)}
    \left(\partial_{\mc E_{x+1}} - \partial_{\mc E_x}\right) e^{-\mc
      U(\mcE)} \gamma^2_{x,{x+1}} \left(\partial_{\mc
        E_{x+1}} - \partial_{\mc E_x}\right)
  \end{split}
\end{equation}
We will use also the finite dimensional generators
\begin{equation}
  \label{eq:genfinite}
  L_k = \sum_{x=-k}^{k-1}  L_{x,x+1}  .
\end{equation}

There is a family of invariant measures given by 
\begin{equation}
  \label{eq:im}
  d\mu_\beta = \prod_x \frac{e^{-U(\mc E_x) -\beta \mc E_x}}{M(\beta)}
  d\mc E_x, \qquad \beta >0 .   
\end{equation}
These probability measure are reversible and the corresponding
dirichlet form are
\begin{equation}
  \label{eq:dirf}
  \begin{split}
    \mc D (f) &= \sum_x \mc D_{x,x+1}(f), \qquad \mc D_{x,x+1}(f) =
    \int \gamma^2_{x, {x+1}} \left( \partial_{\mc E_{x+1}}f
      - \partial_{\mc E_x}f \right)^2 d\mu_\beta\\
    \mc D_k (f) &= \sum_{x=-k}^{k-1} \mc D_{x,x+1}(f) .
  \end{split}
\end{equation}

In this article we will consider only the dynamics in equilibrium,
starting with initial configuration distributed by $\mu_\beta$ for a
given $\beta>0$. From standard arguments it follows, under reasonable
conditions on $\alpha$, the existence of
the solution of the equilibrium dynamics, and that local smooth
functions form a core for the domain of the generator $L$. We will not
worry here about these issues and we assume that all these objects are
well defined.
We denote with $\bP$ the measure on $\cC^0(\bR_+,\bR_+^{\bZ})$ for the
equilibrium dynamics determined by
equation \eqref{eq:sde1} and by $\bE$ the corresponding expectation. We
will also use the notation $\bE(f)=<f>$. 

\section{Equilibrium Fluctuations}
\label{sec:equil-fluct}


Our main goal is to prove the following theorem:
\begin{thmm}
  Let $G(y), F(y)$ smooth function with compact support on $\bb
  R$. Then 
  \begin{equation}
    \label{eq:clt}
    \begin{split}
      \lim_{\ve\to 0} \ve \sum_{x,z} G(\ve x) F(\ve z) \left[\left< \mc
        E_x(\ve^{-2}t) \mc E_z(0)\right> - <\mc E_0>^2 \right] \\
      \ = \chi \iint G(y) F(y') \frac{e^{-(y-y')^2/2tD}}{\sqrt{2\pi t D}} dy\;
      dy'
    \end{split}
  \end{equation}
\end{thmm}
Where $\chi = <\mc E_0^2> - <\mc E_0>^2 = var(\mcE_0)$.
The diffusivity $D$ is given by the usual Green Kubo formula, but will appear
in the proof as a variational formula.

We specify here the assumption under which we are able to prove this
theorem:

\begin{itemize}
\item $\gamma$ is such the following spectral gap bound is satisfied:
  \begin{equation}
    \label{eq:sg}
    \left< f^2 | \bar\mcE_k\right> \le C k^2 \mathcal D_k(f) 
  \end{equation}
for any local function $f$ such that 
$< f^2 | \bar\mcE_k>= 0$, where $<\cdot|  \bar\mcE_k>$ denotes the microcanonical
conditional expectation on the corresponding energy surface 
$\bar\mcE_k = \frac 1{2k+1} \sum_{|x|\le k} \mcE_x$. The constant $C$
is independent of $k$ but can depend on $\bar\mcE_k$.
\item For some $a\ge 1$ there exists $\tgamma(\mcE)$ such that: 
  \begin{equation}
    \label{eq:6}
a^{-1}\tgamma(\mcE_0)\tgamma( \mcE_1) \le \gamma(\mcE_0, \mcE_1) \le
a\tgamma(\mcE_0)\tgamma( \mcE_1). 
  \end{equation}
\end{itemize}

The function $\gamma$ arising in \cite{lojams} satisfies these
conditions (in particular the spectral gap bound is proven in 
\cite{os} with a constant $C$ independent of the energy). It is an
open problem at the moment if the $\gamma$ function arising in 
\cite{dolgoliv} satisfies such conditions.

\section{Time variance}
\label{sec:time-variance}
We start with computing the time evolution of the left hand side of
\eqref{eq:clt} before the limit. 
\begin{equation*}
  \begin{split}
&\ve \sum_{x,z} G(\ve x) F(\ve z) \left\{\left[\left< \mc
        E_x(\ve^{-2}t) \mc E_z(0)\right> - <\mc E_0>^2 \right]
- \left[\left< \mc
        E_x(0) \mc E_z(0)\right> - <\mc E_0>^2 \right]\right\}\\
&    = \ve \sum_{x,z} G(\ve x) F(\ve z) \left<( 
      \mc E_x(\ve^{-2}t) - \mc E_x(0))(\mc E_z(0) -  \bar e)\right> \\
&    = \int_0^{t} \sum_{x,z} [G'(\ve x)+\frac 12G''(\ve x)\ve] F(\ve z) 
    \left<\alpha_{x,x+1}(\ve^{-2}s)(\mc E_z(0) - \bar e) \right> \; ds + \cO(\ve),
  \end{split}
\end{equation*}
where, in the last line, we have used Schwarz inequality,
stationarity, and, since $\alpha_{x,x+1}\in L^2$,  
$$
\|\sum_{x} G'''(\ve z) \alpha_{x,x+1}\|_{L^2}\leq C_\#  \ve^{-1/2}
$$
$$
\|\sum_{z} F(\ve z)(\mc E_z -  \bar e)\|_{L^2}\leq C_\#  \ve^{-1/2}
$$
Note that the term with the second derivative of $G$ has exactly the
same form as the one with $G'$. Thus, given the arbitrariness of $G$,
it suffice to show that the term in $G'$ has limit in order to prove
 that the one with $G''$ vanishes. 
Next choose local smooth functions $f(\mcE)$ (with $<f> = 0$) and set
\begin{equation*}
  \begin{split}
    \phi_f &= \alpha(\mc E_1, \mc E_0) - \kappa (U'(\mc E_1) - U'(\mc
    E_0)) - Lf\\
    &= \alpha(\mc E_1, \mc E_0) - \kappa ([U'(\mc E_1) -\beta] - [U'(\mc
    E_0)- \beta]) - Lf
  \end{split}
\end{equation*}
We will choose later the constant $\kappa = \chi D$ (this is also
called thermal conductivity).
We can then continue our computation:
\begin{align}
     = \int_0^{t} \ve \sum_{x,z} G''(\ve x) F(\ve z) 
    \kappa \left<[U'(\mc E_x(\ve^{-2}s))-\beta](\mc E_z(0) - \bar
      e) \right> \; ds \label{deco1} \\ 
   + \int_0^{t} \sum_{x,z} G'(\ve x) F(\ve z) 
    \left<L \tau_{x}f (\ve^{-2}s)(\mc E_z(0) - \bar e) \right> \; ds \label{deco2}\\
 +\int_0^{t} \sum_{x,z} G'(\ve x) F(\ve z) 
    \left<\tau_{x} \phi_f (\ve^{-2}s)(\mc E_z(0) - \bar e) \right> \; ds
+ o(\ve) \label{deco3}
\end{align}

About line \eqref{deco1}, we prove in section \ref{sec:proof-eqbg} that 
\begin{equation}
  \label{eq:bg}
  \lim_{\ve \to 0} \left< \left( \int_0^t \ve^{1/2} \sum_z G''(\ve z)
      \left[U'(\mc E_x(\ve^{-2}s)) - \beta - \chi^{-1} (\mc
        E_x(\ve^{-2}s)- \bar e) \right] ds  \right)^2\right> = 0
\end{equation}

Since $D= \kappa/\chi$, 
\begin{equation*}
  \begin{split}
    \left| \int_0^{t} \ve \sum_{x,z} G''(\ve x) F(\ve z) \left<\left\{\kappa
        [U'(\mc E_x(\ve^{-2}s)) -\beta] - D [(\mc E_x(\ve^{-2}s))-\bar
        e]\right\}(\mc E_z(0) - \bar e) \right> \; ds\right|^2\\
    \le \left<\left(\ve^{1/2}\sum_z F(\ve z) (\mc E_z(0) - \bar
        e)\right)^2\right> \\
    \cdot\left< \left( \int_0^t\ve^{1/2} \sum_z G''(\ve z)
        \kappa\left[U'(\mc E_x(\ve^{-2}s)) - \beta - \chi^{-1} (\mc
          E_x(\ve^{-2}s)- \bar e) \right] ds \right)^2\right>
  \end{split}
\end{equation*}
This will close our equation if we prove that
\eqref{deco2} and \eqref{deco3} lines will converge to 0 
after the limit as $\ve \to 0$ and minimization on the local function
$f$.

Second term is easy:
\begin{equation*}
  \begin{split}
    \left|\int_0^{t} \ve^2\sum_{x,z} G'(\ve x) F(\ve z) \left<\ve^{-2} L
      \tau_{x}f (\ve^{-2}s)(\mc E_z(0) - \bar e) \right> \; ds\right|^2\\
   =\left|(\mc E_z(0) - \bar e) \ve^2\sum_{x,z} G'(\ve x) F(\ve z) \left< 
      (\tau_{x}f (\ve^{-2}t) - \tau_{x}f (0)) (\mc E_z(0) - \bar e)
    \right>\right|^2\\
    \le \ve^2 \left<\left(\ve^{1/2}\sum_z F(\ve z) (\mc E_z(0) - \bar
        e)\right)^2\right>
    \left<\left(\ve^{1/2}\sum_x G'(\ve x)  (\tau_{x}f (\ve^{-2}t) -
        \tau_{x} f (0)) \right)^2\right>
  \end{split}
\end{equation*}
since
\begin{equation*}
  \left<\left(\ve^{1/2}\sum_z F(\ve z) (\mc E_z(0) - \bar
        e)\right)^2\right> = \ve \sum_z F(\ve z)^2 \chi \le \chi \|F\|_2^2
\end{equation*}
and similarly 
\begin{equation*}
   \left<\left(\ve^{1/2}\sum_x G'(\ve x)  (\tau_{x}f (\ve^{-2}t) -
        \tau_{x}f (0)) \right)^2\right> \le C\|G'\|^2 \sum_x <\tau_x
    f, f>\ < \ \infty
\end{equation*}
since $f$ is local and of null average.

About the third term, using again Schwarz inequality, the square is
bounded by
\begin{equation*}
  \chi \|F\|_2^2 \left< \left(\int_0^{t} \ve^{-1/2} \sum_{x} G'(\ve x) 
    \tau_{x} \phi_f(\ve^{-2}s)ds\right)^2 \right>
\end{equation*}
The rest of the work is in order to prove that
\begin{equation*}
  \inf_f \lim_{\ve\to 0} \left< \left(\int_0^{t} \ve^{-1/2} \sum_{x} G'(\ve x) 
    \tau_{x} \phi_f(\ve^{-2}s)ds\right)^2 \right> = 0
\end{equation*}

By the time variance estimate for a stationary markov process 
(see (4.1) in \cite{os} or in chapter 2 of \cite{klo}):
\begin{equation*}
  \begin{split}
    &\left< \left(\int_0^{t} \ve^{-1/2}\sum_{x} G'(\ve x) \tau_{x}
        \phi_f(\ve^{-2}s)ds\right)^2 \right> \\
    &\le 16 t
    \left<\left(\ve^{1/2}\sum_{x} G'(\ve x) \tau_{x} \phi_f \right) (-\ve^{-2}L)^{-1}
      \left(\ve^{1/2}\sum_{x} G'(\ve x) \tau_{x} \phi_f\right)\right> \\
    &= 16 t \sup_g \left\{\ve^{1/2} \sum_x G'(\ve x) <\tau_x \phi_f
      g> - \ve^{-2}\mc D(g)\right\}
  \end{split}
\end{equation*}

After some steps (see proof of theorem 2 in \cite{os}) we obtain
that, for $k << \ve^{-1}$ this is bounded by
\begin{equation*}
  Ct \ve \sum_x G'(\ve x)^2 (2k+1) \left< Av_{k'}(\phi_f),
    (-L_k)^{-1} Av_{k'}(\phi_f)\right> - C_G O(\ve k)
\end{equation*}
where $Av_k(\phi) = \frac 1{2k+1} \sum_{j=-k}^k \tau_j \phi$, and $k'
< k$ such that $Av_k(\phi)$ is localized between $(-k, \dots, k)$.

Now comes the hard work.
Define the vector fields 
$$
\partial_{x,x+1} = \gamma_{x, {x+1}} \left(\partial_{\mc
  E_{x+1}}- \partial_{\mc E_{x}}\right)
$$
 and 
$$\partial_{x,x+1}^* = e^{\mathcal U} \left(\partial_{\mc
  E_{x+1}}- \partial_{\mc E_{x}}\right) e^{-\mathcal U}\gamma_{x,x+1}
$$ 
its adjoint with respect to $d\mu$. Then we car write $L_{x,x+1}
= \partial_{x,x+1}^* \partial_{x,x+1}$.

For any $k\in \mathbb N_*$, define the microcanonical expectation 
$\mc M_k \phi = \left< \phi |\bar{\mc E}_k\right>$
 where $\bar{\mc E}_k = (2k+1)^{-1}\sum_{|x|\le k} \mc E_x$.
Define the space of functions on $\bb R_+^{\bb Z}$
\begin{equation}
  \label{eq:C0}
  \mc C_0 = \left\{ \phi \in L^2(\mu) \ \text{local} : \;\mc M_{k_0}
    \phi   = 0\ \text{for a finite}\; k_0\right\} 
\end{equation}

Since $L_{k_0}$, on each microcanonical surface, has a spectral gap
bounded below by $k_0^{-2}$ uniformly in the energy\footnote{Notice that
  it is not necessary here a uniform bound, since $k_0$ is fixed.},
we can invert $u_{k_0} = (-L_{k_0})^{-1} \phi$, for $\phi \in \mc C_0$
and we have
\begin{equation*}
   \mc D_{k_0} (u_{k_0}) =\ <u_{k_0}, \phi>\ \le\ <u_{k_0}^2>^{1/2} <\phi^2>^{1/2} \le C^{1/2}
   k_0 \mc D_{k_0} (u_{k_0})^{1/2}  <\phi^2>^{1/2}
\end{equation*}
and consequently 
\begin{equation}
  \label{eq:ukb}
  \mc D_{k_0} (u_{k_0}) \le C k_0^2 <\phi^2>
\end{equation}
In particular for any smooth local function $h$:
\begin{equation}
  \label{eq:ukb2}
  \left< \phi, h \right> = \sum_{i=-k_0}^{k_0-1} \left< (\partial_{i,i+1} u_{k_0}), (\partial_{i,i+1} h)
  \right> \le C^{1/2} k_0  <\phi^2>^{1/2} \mc D_{k_0} (h)^{1/2}
\end{equation}



\medskip

For a local function $\phi\in \mc C_0$, we will show that the
following limit exists 
\begin{equation}
  \label{eq:triple}
  \lim_{k\to\infty} (2k+1) \left< Av_{k'}(\phi),
    (-L_k)^{-1} Av_{k'}(\phi)\right> =  \triple \phi\triple_{-1}^2
\end{equation}
we will compute such norm and to conclude the proof we need to show that
\begin{equation}
  \label{eq:2}
  \inf_f \triple \phi_f \triple_{-1} = 0.
\end{equation}

\section{Variational formula for the limit space-time variance}
\label{sec:vari-form-limit}

Let us compute the limit \eqref{eq:triple}. For simplicity of notations, assume
that $k_0 = 0$, i.e. $\phi= \partial_{0,1}^* F$, with $F = X_{0,1}
u_0$ (following strictly the notation of the previous section). The
general case, $k_0 < \infty$, follows easily by linearity.
 
\begin{equation*}
  \begin{split}
    &(2k+1) \left< Av_{k'}(\phi), (-L_k)^{-1} Av_{k'}(\phi)\right> \\
    &= \sup_h \left\{2 \left<\phi, \frac 1{2k+1} \sum_{|j|\le k'}\tau_j
        h\right> - \frac 1{2k+1} \mc D_k(h) \right\}\\
    &= \sup_h \left\{2 \left<F, \partial_{0,1}(\frac 1{2k+1} \sum_{|j|\le k'}\tau_j
        h)\right> - \frac 1{2k+1} \mc D_k(h) \right\}\\
  \end{split}
\end{equation*}
Call $\xi^k(h) = \partial_{0,1}(\frac 1{2k+1} \sum_{|j|\le k'}\tau_j
        h)$, and observe that
\begin{equation*}
  \left<F, \xi^k(h)\right> \le \frac{C\|F\|_2}{2k} \mc D_k(h)^{1/2} 
\end{equation*}
and that by Schwarz inequality
\begin{equation*}
  <\xi^k(h)^2> \le \frac{1}{2k+1} \sum_{|j|\le k'} \left<(\partial_{0,1}
  \tau_j h)^2\right> \le \frac 1{2k}\mc D_k(h)
\end{equation*}
So we obtain the upper bound
\begin{equation*}
  \begin{split}
    (2k+1) \left< Av_{k'}(\phi), (-L_k)^{-1} Av_{k'}(\phi)\right> \\
    \le  \sup_h \left\{ 2 <F, \xi^k(h)> -    <\xi^k(h)^2> \right\}
  \end{split}
\end{equation*}
Since we can restrict the supremum on functions $h$ such that \newline
$<\xi^k(h)^2> \le <F^2>$, for any of such function $h$ we can extract
convergent subsequences in $L^2(d\mu)$. 

Observe that 
\begin{equation}
\partial_{z,z+1} \tau_x \xi^k(h) = \partial_{x,x+1}
\tau_z \xi^k(h)\label{eq:closedf}
\end{equation}
as long as $|x|$ and $|z|$ are small with respect
to $k$ and $|x-z|\ge 2$,
(for $|x-z|<2$ there are some relations that
we will have to take into account). So any limit $\xi(h)$ for
$k\to\infty$ enjoy of this 
property for any $x,z \in \bb Z$. We call \emph{closed
  forms} (or germs of closed forms) all functions that satisfy
property \eqref{eq:closedf},
and we denote the closed subset of
such functions in $L^2(\mu)$ by $\frak h_c$. So we have proved that, 
for $\phi= \partial_{0,1}^* F$, with $F = X_{0,1} u_0$ :
\begin{equation}
  \label{eq:1}
  \lim_{k\to\infty} (2k+1) \left< Av_{k'}(\phi), (-L_k)^{-1}
    Av_{k'}(\phi)\right> 
  \le \sup_{\xi\in \frak h_c} \left\{ 2 <F, \xi> -    <\xi^2> \right\}
\end{equation}

\bigskip

Let us now study a lower bound.
Observe that $L_k \sum_{|x|\le k} x \mc E_x = \sum_{x=-k}^{k-1} \tau_x
\alpha_{0,1}$. 
Computing we have:
\begin{equation}
  \label{eq:var1}
  (2k+1) \left< Av_k\alpha_{0,1} , (-L_k)^{-1}  Av_k \phi\right> \
  \mathop{\longrightarrow}_{k\to\infty} \ -<\gamma_{0,1} F>
\end{equation}
For a local smooth function $f$
\begin{equation}
  \label{eq:var2}
  (2k+1) \left< Av_{k'} Lf , (-L_k)^{-1}  Av_k \phi\right>
  \mathop{\longrightarrow}_{k\to\infty} \ 
  - \sum_x <\phi \tau_x f> = - <F\partial_{0,1} \Gamma_f> 
\end{equation}
where we have defined the formal sum $\Gamma_f = \sum_x \tau_x f$
(since $f$ is local, $\partial_{0,1} \Gamma_f$ is well define finite sum).
Similarly:
\begin{equation}
  \label{eq:var3}
   (2k+1) \left< Av_{k'} Lf , (-L_k)^{-1}  Av_{k'} Lf\right>
 \mathop{\longrightarrow}_{k\to\infty} \ 
 <(\partial_{0,1} \Gamma_f)^2>
\end{equation}
and all together, for any $a\in \bb R$:
\begin{equation}
  \label{eq:var2}
  \begin{split}
    (2k+1) \left< Av_{k'} (a\alpha_{0,1} + Lf) , (-L_k)^{-1} Av_{k'}
    (a\alpha_{0,1} + Lf) \right>\\
    \mathop{\longrightarrow}_{k\to\infty} \ 
    <(a\gamma_{0,1} + \partial_{0,1}\Gamma_f)^2>
  \end{split}
\end{equation}
Then for any  $a\in \bb R$ and local $f$:
\begin{equation*}
  \begin{split}
      (2k+1) &\left< Av_{k'}(\phi), (-L_k)^{-1} Av_{k'}(\phi)\right> \\
      &\ge   2 (2k+1) \left< Av_{k'}(\phi), (-L_k)^{-1}
        Av_{k'}(a\alpha_{0,1} + Lf)\right>\\
     & -  (2k+1) \left< Av_{k'} (a\alpha_{0,1} + Lf) , (-L_k)^{-1} Av_{k'}
    (a\alpha_{0,1} + Lf) \right>\\
   &\mathop{\longrightarrow}_{k\to\infty} \
   2a <\gamma_{0,1} F> + 2 <F\partial_{0,1} \Gamma_f> -
   <(a\gamma_{0,1} + \partial_{0,1}\Gamma_f)^2> 
  \end{split}
\end{equation*}
So we have obtained 
\begin{equation}
  \label{eq:lbtriple}
  \triple \phi\triple^2_{-1} \ge \sup_{a,f} \left\{2 <F,a\gamma_{0,1}
    + \partial_{0,1} \Gamma_f> - 
   <(a\gamma_{0,1} + \partial_{0,1}\Gamma_f)^2> \right\}
\end{equation}
In order to show equality in this formula,
 we have to put together the upper and lower
bound, i.e. to prove that every \emph{closed form} $\xi\in\frak h_c$
can be approximated in $L^2(\mu)$ by functions of the type
 $a\gamma_{0,1} + \partial_{0,1}\Gamma_f$, with local f, that we call
 \emph{exact forms}. This is usually the hardest part of the proof,
 where a uniform bound on the spectral gap is needed.
We prove this in section \autoref{sec:closed-forms}.

\bigskip

The general case, for $k_0 > 0$ is obtained in the same way and 
for $\phi = -L_{k_0} u_{k_0}$ we have
\begin{equation}
  \label{eq:var}
  \triple \phi\triple^2_{-1} = \sup_{a,f}
  \left\{2\sum_{i=-k_0}^{k_0} <\partial_{i,i+1} u_{k_0} , \tau_{i}\left(a\gamma_{0,1}
    + \partial_{0,1} \Gamma_f\right)> - 
   <(a\gamma_{0,1} + \partial_{0,1}\Gamma_f)^2> \right\}
\end{equation}

\section{Hilbert Space of Fluctuations}
\label{sec:hilb-space-fluct}

By polarizing the $\triple\cdot\triple_{-1}$ norm, we can define a
scalar product that we denote by $<<\cdot,\cdot>>_{-1}$, and the
corresponding Hilbert Space by $\mc H_{-1}$. 
It is also clear, prom the results of the previous section that
\begin{equation}
  \label{eq:scp}
  <<\phi,\psi>>_{-1} = \lim_{k\to\infty} (2k+1) \left< Av_{k'}(\phi),
    (-L_k)^{-1} Av_{k'}(\psi)\right> 
\end{equation}
Straightforward calulations show that, denoting $\nabla_{0,1} U' =
U'(\mc E_1) - U'(\mc E_0)$,  
\begin{equation}
  \label{eq:ortho}
  <<\nabla_{0,1} U', Lf>>_{-1} = 0
\end{equation}
and we have computed already, for $\phi = \partial^*_{0,1} F$:
\begin{equation}
  \label{eq:expl}
  \begin{split}
    <<\alpha_{0,1},\psi>>_{-1} = - <\gamma_{0,1} F>\\
     <<Lf,\psi>>_{-1} = - <\partial_{0,1}\Gamma_f, F>\\
  \end{split}
\end{equation}
in particular
\begin{equation}
  \label{eq:expl1}
  <<\alpha_{0,1},\nabla_{0,1} U'>>_{-1} = -1
\end{equation}
and
\begin{equation}
  \label{eq:3}
  \triple \alpha_{0,1}\triple_{-1}^2 = <\gamma_{0,1}^2>
\end{equation}

\begin{prop}
  \begin{equation}
    \label{eq:decomp}
    \mc H_{-1} = \text{Clos}{\{Lf, f\text{local smooth}\}} \plus
    {\{a\alpha_{0,1}, a\in \bb R\}}
  \end{equation}
\begin{equation}
    \label{eq:decomport}
    \mc H_{-1} = \text{Clos}{\{Lf, f\text{local smooth}\}} \oplus
    {\{a\nabla_{0,1} U', a\in \bb R\}}
  \end{equation}
\end{prop}

This proposition assure that $\inf_f \triple \phi_f\triple_{-1} = 0$.

\begin{proof}
  Consider first $\phi = \partial^*_{0,1} F$. Then as consequence of
  all above we have the variational formula:
  \begin{equation*}
    \triple \phi\triple_{-1}^2 = \sup_{a\in\mathbb R, g\ \text{loc}} 
    \left\{ 2<F, a\gamma_{0,1} + \partial_{0,1}\Gamma_g> - <(
      a\gamma_{0,1} + \partial_{0,1}\Gamma_g)^2> \right].
  \end{equation*}
In particular for $\phi_f$ we have $F = \gamma_{0,1} - \kappa \gamma_{0,1}^{-1}
+ \partial_{0,1}\Gamma_f$.
 
 Observe that $<\gamma_{0,1}^{-1},\partial_{0,1}\Gamma_g> = 0$. 
So we have
\begin{equation*}
  \begin{split}
    \triple \phi_f\triple_{-1}^2 = \sup_{a\in\mathbb R, g\ \text{loc}}
    \left\{-2a\kappa + 2<\gamma_{0,1} + \partial_{0,1}\Gamma_f ,
      a\gamma_{0,1} + \partial_{0,1}\Gamma_g> - <( a\gamma_{0,1}
      + \partial_{0,1}\Gamma_g)^2> \right\}\\
    = \sup_{a\in\mathbb R, g\ \text{loc}}
    \left\{-2a\kappa + 2a<\gamma_{0,1} + \partial_{0,1}\Gamma_f ,
      \gamma_{0,1} + \partial_{0,1}\Gamma_g> - a^2<( \gamma_{0,1}
      + \partial_{0,1}\Gamma_g)^2> \right\}\\
    = \sup_{a\in\mathbb R}
    \left\{-2a\kappa + (2a - a^2)<( \gamma_{0,1}
      + \partial_{0,1}\Gamma_f)^2> \right\}\\
    = \frac{<\left( \gamma_{0,1}
      + \partial_{0,1}\Gamma_f)^2> - \kappa\right)^2>}{<( \gamma_{0,1}
      + \partial_{0,1}\Gamma_f)^2>}.
  \end{split}
\end{equation*}
Defining
\begin{equation}
  \label{eq:7}
  \kappa = \inf_f <(\gamma_{0,1} + \partial_{0,1}\Gamma_f)^2>
\end{equation}
we obtain the result.
\end{proof}

\section{Closed forms}
\label{sec:closed-forms}

Recall $\mf h_c \subset L^2(\mu)$ is the space of the function $\xi$
such that
\begin{equation}
  \label{eq:closed rel2}
  \begin{split}
    \partial_{x,x+1} \tau_y \xi &= \partial_{y,y+1} \tau_x \xi, \qquad
    |x-y|\ge 2\\
    \partial_{x,x+1} \tau_{x+1} \xi &= \partial_{x+1,x+2} \tau_x \xi +
     \left(\partial_{x,x+1} \log \gamma_{x+1,x+2}\right) \tau_{x+1} \xi
     - \left(\partial_{x+1,x+2}\log\gamma_{x,x+1}\right) \tau_x \xi
  \end{split}
\end{equation}
For any local smooth f, we call \emph{exact forms} functions of the type
\begin{equation}
  \label{eq:exact}
  \partial_{0,1} \Gamma_f
\end{equation}
Remark that satisfy \eqref{eq:closed rel2} (i.e. they are closed).

Also the function $\xi = \gamma_{0,1}$ satisfy \eqref{eq:closed rel2}.
More generally we call \emph{exact} the function of the form $a\gamma_{0,1}
+ \partial_{0,1}\Gamma_f$. We want to prove that $\mf h_c$ is the
closure in $L^2$ of such exact functions.

Strategy:

We take $\xi\in \mf h_c$ and define for $y=-k, \dots,k$, 
$$
\xi^{(k)}_y = \mathbb E\left(\tau_y\xi \big| \mc
  F_k\right)\varphi(\bar{\mc E_k}), \qquad \mc
F_k = \sigma\left(\mc E_x, |x|\le k
\right) 
$$ 
where $\varphi$ is a smooth function on $\mathbb R_+$ with compact
support, such that  $\varphi(E(\beta)) = 1$ and it is bounded by
1. This is a cutoff function that we need in order to make uniform
bounds later.

Now we are in the classical finite dimensional situation, and on
each microcanonical surface ($\bar{\mc E_k} = e$ fixed)
$\left\{\xi^{(k)}_y, y=-k, \dots,k\right\}$ is a closed form (in a
weak sense), i.e. exact, so it can be integrated. 
Let $u_k ( \mc E_x, |x|\le k, \bar{\mc E_k})$
  such that $\partial_{y,y+1} u_k = \xi^{(k)}_y$ for $y= -k, \dots,k-1$.
This derivative should be intended in the distributional sense. 
We can always recenter $u_k$ in such a way that 
$\mathbb E\left(u_k \big| \mc E_x, |x|\le k \right) = 0$.

\bigskip

By condition \eqref{eq:6}, we can directly assume that 
 $\gamma(\mcE_0, \mcE_1) =
\tgamma(\mcE_0)\tgamma( \mcE_1)$, since the corresponding dirichlet
forms are equivalent, and consequently all approximations done in the
corresponding Hilbert spaces.

Then we define 
\begin{equation*}
  u_{k,l} = \frac 1{2(k+l)<\gamma^2>^2} \mathbb E\left(
    \tgamma_{k+l+1}^2 \tgamma_{-k-l-1}^2 u_{2k} \big| \mc F_{k+l}\right)
\end{equation*}
\begin{equation*}
  \widehat u^{(k)} = \frac 4k \sum_{l=k/2}^{3k/4}  u_{k,l}
\end{equation*}
Now we compute 
\begin{equation*}
  \partial_{0,1} \Gamma_{ \widehat u^{(k)}} = \partial_{0,1}
    \sum_{x=-\infty}^{+\infty} \tau_x \widehat u^{(k)}
\end{equation*}

\bigskip
In the calculation there are \emph{interior} terms that depends on
$$
\partial_{x,x+1} \widehat u^{(k)} 
$$
that converges in $L^2(\mu)$ to $\tau_x \xi$.

Then there are boundary terms more tricky to control that, for $x$
enough small, will
converge to $a\tgamma_{0}\tgamma_1$ for some $a\in \mathbb R$.

One of the boundary terms is given by 
\begin{equation}\label{eq:boundary}
  \begin{split}
   \frac 4k \sum_{l=k/2}^{3k/4}  \frac 1{2(k+l)<\gamma^2>}
   \tau_{-k-l}\left\{\tgamma_{k+l}\tgamma_{k+l+1} 
      \mathbb E\left( \tgamma_{k+l+1}^2 \tgamma_{-k-l-1}^2
        \partial_{\mc E_{k+l}} u_{2k} \big| \mc
        F_{k+l}\right)\right\}\\
    = -\frac 4k \sum_{l=k/2}^{3k/4}  \frac 1{2(k+l)<\gamma^2>}
    \tgamma_{1} \tau_{-k-l} \mathbb E\left( 
        \tgamma_{-k-l-1}^2\tgamma_{k+l+1} \partial_{k+l+1,k+l} u_{2k} \big| \mc
        F_{k+l}\right)\\
      +  \frac 4k \sum_{l=k/2}^{3k/4}  \frac 1{2(k+l)<\gamma^2>}
    \tgamma_{1} \tau_{-k-l} \mathbb E\left( \tgamma_{k+l} \tgamma_{k+l+1}^2
        \tgamma_{-k-l-1}^2 \partial_{\mc E_{k+l+1}} u_{2k} \big| \mc
        F_{k+l}\right)\\
  \end{split}
\end{equation}
the other boundary terms are similar and will be treated in the same
way.
The first term of the rhs of \eqref{eq:boundary} converges to $0$ in
$L^2$, because  $\partial_{k+l+1,k+l} u_{2k}$ are uniformly bounded in
$L^2$.
The second term of the rhs of  \eqref{eq:boundary}, after an
integration by part, became
\begin{equation}\label{eq:boundary2}
  \begin{split}
    \frac 4k \sum_{l=k/2}^{3k/4} \frac 1{2(k+l)<\gamma^2>}
     \tgamma_1 \tgamma_0 \tau_{-k-l} \mathbb E\left( (\partial_{\mc
         E_{k+l+1}}^*\tgamma_{k+l+1}^2) 
        \tgamma_{-k-l-1}^2\ u_{2k} \big| \mc
        F_{k+l}\right) \\
      = \tgamma_{0}\tgamma_1 \ h_k(\mc E_0, \mc E_{-1},\dots, \mc E_{-7k/2} )
  \end{split}
\end{equation}
where
\begin{equation*}
  \partial_{\mc E_{k+l+1}}^*\tgamma_{k+l+1}^2 = 
  - \partial_{\mc E_{k+l+1}}\tgamma_{k+l,k+l+1}^2 + U'({\mc E_{k+l+1}})
  \tgamma_{k+l,k+l+1}^2
\end{equation*}


We rewrite $h_k$ as 

\begin{equation}\label{eq:boundary3}
  \begin{split}
   <\tgamma^2>^{2} h_k =\frac 4k \sum_{l=k/2}^{3k/4}  \frac 1{2(k+l)}
    \tau_{-k-l} \mathbb E\left( (\partial_{\mc E_{k+l+1}}^*\tgamma_{k+l+1}^2)
        \tgamma_{-k-l-1}^2\ u_{2k} \big| \mc
        F_{k+l}\right)\\
      = \frac 4k \sum_{l=k/2}^{3k/4}  \frac {1}{2(k+l)}
    \tau_{-k-l} \left[\mathbb E\left( (\partial_{\mc
          E_{k+l+1}}^*\tgamma_{k+l+1}^2) 
        \tgamma_{-k-l-1}^2\ u_{2k} \big| \mc F_{k+l}\right)\right]\\
 = \frac 4k  \sum_{l=k/2}^{3k/4}  \frac {1}{2(k+l)}
    \tau_{-k-l} \mathbb E\left( (\partial_{\mc E_{k+l+1}}^*\tgamma_{k+l+1}^2) 
        \tgamma_{-k-l-1}^2\ u_{2k} \big| \mc F_{k+l}\right)\\
= \frac 4k  \sum_{l=k/2}^{3k/4}  
   \frac{1}{2(k+l)} \tau_{-k-l} \mathbb E\left(
      \frac{\tgamma_{-k-l-1}^2}{k-l}\sum_{j=k+l+1}^{2k} 
      \partial_{\mc  E_{j}}^*\tgamma_{j}^2 (u_{2k} \circ \pi^{j,k+l+1}) 
        \big| \mc F_{k+l}\right)\\
= \frac 4k \sum_{l=k/2}^{3k/4}  
   \frac{1}{2(k+l)} \tau_{-k-l} \mathbb E\left(
       u_{2k}\frac{\tgamma_{-k-l-1}^2}{k-l}\sum_{j=k+l+1}^{2k} 
\partial_{\mc E_{j}}^*\tgamma_{j}^2 \big| \mc F_{k+l}\right)\\
 +\frac 4k \sum_{l=k/2}^{3k/4}  
   \frac{1}{2(k+l)} \tau_{-k-l} \mathbb E\left(
      \frac{\tgamma_{-k-l-1}^2}{k-l}\sum_{j=k+l+1}^{2k} \partial_{\mc
          E_{j}}^*\tgamma_{j}^2 (u_{2k} \circ \pi^{j,k+l+1} - u_{2k}) 
        \big| \mc F_{k+l}\right)\\
  \end{split}
\end{equation}

Let us estimate the first expression of the rhs, by Schwarz
inequality and the cutoff introduced, its square is bounded by
\begin{equation*}
  \begin{split}
   \tau_{-k-l} \mathbb E\left( u_{2k}^2 \frac 4k \sum_{l=k/2}^{3k/4}
      \tgamma_{-k-l-1}^2 \Big| \mc F_{k+l}\right) \left(\frac 4k
    \sum_{l=k/2}^{3k/4}
    \frac{\tgamma_{-k-l-1}^2}{4(k+l)^2} 
  \tau_{-k-l}\mathbb E\left(\left[\frac 1{k-l}\sum_{j=k+l+1}^{2k} 
      \partial_{\mc E_{j}}^*\tgamma_{j}\right]^2 \big| \mc
    F_{k+l}\right)\right)\\
= \tau_{-k-l} \mathbb E\left( u_{2k}^2 \frac 4k \sum_{l=k/2}^{3k/4}
      \tgamma_{-k-l-1}^2 \Big| \mc F_{k+l}\right) \left(\frac 4k
    \sum_{l=k/2}^{3k/4}
    \frac{\tgamma_{-k-l-1}^2}{4(k+l)^2} 
  \frac 1{(k-l)^2}\sum_{j=k+l+1}^{2k} 
    \tau_{-k-l} \mathbb E\left( (\partial_{\mc E_{j}}^*\tgamma_{j})^2 \big| \mc
    F_{k+l}\right)\right)\\
\le C \tau_{-k-l} \mathbb E\left( u_{2k}^2 \Big| \mc F_{k+l}\right) \left(\frac 4k
    \sum_{l=k/2}^{3k/4}
    \frac{\tgamma_{-k-l-1}^2}{4(k+l)^2(k-l)} \right)
\le  \frac {C'}{k^3} \tau_{-k-l} \mathbb E\left( u_{2k}^2
  \Big| \mc F_{k+l}\right) 
  \end{split}
\end{equation*}
Since by the spectral gap $<u_{2k}^2> \le Ck^3$ the $\mu$ expectation
of this term is bounded in $k$.

Also the estimate of the second term follows as in \cite{os}. 
The expectation of the square of the second term of
\eqref{eq:boundary3} is bounded by 
\begin{equation*}
  \begin{split}
    \left<\frac 4k \sum_{l=k/2}^{3k/4}  
   \frac{1}{4(k+l)^2} \tgamma_{-2k-2l-1}^4 \right> 
 \left<\frac 4k \sum_{l=k/2}^{3k/4}
   \frac{1}{(k-l)^2}\tau_{-k-l} \mathbb  E\left(\sum_{j=k+l+1}^{2k} 
      \partial_{\mc E_{j}}^*\tgamma_{j}^2 (u_{2k} \circ \pi^{j,k+l+1} - u_{2k}) 
        \big| \mc F_{k+l}\right)^2 \right>\\
    \le \frac{C}{k^2} \frac 4k \sum_{l=k/2}^{3k/4}
   \frac{1}{(k-l)^2} \left<\left(\sum_{j=k+l+1}^{2k} 
      \partial_{\mc E_{j}}^*\tgamma_{j}^2 (u_{2k} \circ \pi^{j,k+l+1} - u_{2k}) 
        \right)^2 \right>\\
      \le \frac{C}{k^2} \frac 4k \sum_{l=k/2}^{3k/4}
   \frac{1}{k-l} \sum_{j=k+l+1}^{2k} 
   \left<\left(u_{2k} \circ \pi^{j,k+l+1} - u_{2k}
 \right)^2 \right>\\
= \frac{C}{k^2} \frac 4k \sum_{l=k/2}^{3k/4}
   \frac{1}{k-l} \sum_{j=k+l+1}^{2k} 
   \sum_{i=j}^{k+l} (k+l-j) \left<( u_{2k} \circ \pi^{i,i+1} - u_{2k})^2 \right>
  \end{split}
\end{equation*}
 
Then we can bound
\begin{equation*}
  \left<( u_{2k} \circ \pi^{i,i+1} - u_{2k})^2 \right> \le \mathcal D_i(u_{2k})
\end{equation*}
following the same argument as in \cite{os}, concluding the proof.

\section{Proof of Boltzmann-Gibbs principle}
\label{sec:proof-eqbg}

We prove here \eqref{eq:bg}.

Let us denote
\begin{equation*}
  \phi_x = U'(\mc E_x) - \beta - \chi^{-1} (\mc E_x- \bar e)
\end{equation*}
Notice that this function is not in $\mc C_0$. But it has the
following property. Define the microcanonical expectation $\Gamma_k
\phi_0 = \left< \phi_0 |\bar{\mc E}_k\right>$
 where $\bar{\mc E}_k = (2k+1)^{-1}\sum_{|x|\le k} \mc E_x$.

This actually can be explicitly computed and is gives
\begin{equation}
  \label{eq:expl}
   \left<U'(\mc E_0)| \bar{\mc E}_k \right> = \beta + \frac 1{2k+1} \partial_e
  \log f_k(\bar{\mc E}_k) 
\end{equation}
where $f_k(e)$ is the probability density of $\bar{\mc E}_k$ under
$d\mu_\beta$. Using using theorem 1.4.1 of my large deviation course
(consequence of the local CLT) we have that 
\begin{equation*}
  \log f_k(e) = I(e) + \log g_k(e') 
\end{equation*}
where $g_k$ has order $\sqrt k$. Since $\partial_e I(e) = \chi (e-\bar
e) + O(e - \bar e)^2$, we have that 
\begin{equation*}
  \Gamma_k(\phi_0)(\bar{\mc E}_k) = O(\bar{\mc E}_k - \bar e)^2 + O(k^{-1})
\end{equation*}
Since $<(\bar{\mc E}_k - \bar e)^4> = O(k^{-2})$ we have consequently: 
\begin{lem}
  \begin{equation}
  k^2\left< (\Gamma_k\phi_0)^2 \right> \le C \label{eq:eqens}
\end{equation}
\end{lem}

This is true because of the \emph{first order} correction in $\phi_0$,
one can see easily that for the function $\mc E_0- \bar e$ this is false.
 
Define now $\tilde \phi_x = \phi_x - \tau_x \Gamma_k\phi_0 = \tau_x
\tilde\phi_0$. Then
\begin{equation*}
  \begin{split}
    \left< \left( \int_0^t \ve^{1/2} \sum_z G''(\ve z)
        \phi_z(\ve^{-2}s))\; ds
      \right)^2\right> \\
    \le
2 \left< \left( \int_0^t\ve^{1/2} \sum_z G''(\ve z)
    \tilde\phi_z(\ve^{-2}s))\; ds
      \right)^2\right> \\ 
    +2 t^2 \left< \left( \ve^{1/2} \sum_z G''(\ve z) \tau_z\Gamma_k\phi_0)
      \right)^2\right>
  \end{split}
\end{equation*}
Since $\tilde \phi_0$ is centered in every microcanonical, we can
solve the equation $L_k u_k = \tilde\phi_0$, i.e. $\tilde\phi \in \mc
C_0$ and the first term goes to $0$ as $\ve \to 0$ by the same
estimates used in section \ref{sec:vari-form-limit}.

Computing the last term we have:
\begin{equation*}
  \begin{split}
   2t^2 \ve \sum_{z,x} G''(\ve z) G''(\ve x)
   <\tau_{z-x}\Gamma_k\phi_0, \Gamma_k\phi_0>\\
   \le t^2 \ve \sum_{z,x} \left(G''(\ve z)^2+ G''(\ve x)^2\right)
   <\tau_{z-x}\Gamma_k\phi_0, \Gamma_k\phi_0>\\
   = 2t^2 \ve \sum_{z} G''(\ve z)^2
   \sum_x <\tau_{x}\Gamma_k\phi_0, \Gamma_k\phi_0>\\
   \le C t^2 \ve \sum_{z} G''(\ve z)^2
   k <(\Gamma_k\phi_0)^2>
\le C t^2 \|G''\|_2^2  k <(\Gamma_k\phi_0)^2>
  \end{split}
\end{equation*}
that goes to $0$ as $k\to\infty$.


\begin{thebibliography}{999}
\footnotesize





\bibitem{BLR} F. Bonetto, J.L. Lebowitz, Rey-Bellet, {\em Fourier's
    law: A challenge to theorists}, Mathematical Physics 2000,
  Imperial College Press, London, 2000, pp.128-150. 

\bibitem{cerrai} S. Cerrai, Ph. Cl\'ement, \emph{Well-posedness of the
  martingale problem for some degenerate diffusion processes occurring
  in dynamics of populations}, Bull. Sci. Math.128 (2004) 355--389. 

\bibitem{dolgoliv} Dmitry Dolgopyat, Carlangelo Liverani, \emph{Energy
  transfer in a fast-slow Hamiltonian system}, Communications in
  Mathematical Physics, \textbf{308}, N. 1, 201-225 (2011). 




 \bibitem{GPV} Guo, M. Z.; Papanicolaou, G. C.; Varadhan, S. R. S.,
   {\em  Nonlinear diffusion limit for a system with nearest neighbor
     interactions}.  Comm. Math. Phys.  {\bf 118}  (1988),  no. 1,
   31--59. 



 \bibitem{klo} T. Komorowski, C. Landim, S. Olla, \emph{Fluctuations in
     Markov Processes},  Springer, to appear (2012). 

\bibitem{lojams} C. Liverani, S. Olla, \emph{Toward the Fourier law
    for a weakly interacting anharmonic crystal}, JAMS 25, {\bf 2},
  April 2012, 555-583. 

\bibitem{os} S. Olla, M. Sasada, \emph{Macroscopic energy diffusion
    for a chain of anharmonic oscillators}, Probability Theory and
  Related Fields. {\bf157} (2013), 721-775. 

\bibitem {ovy} S. Olla, S. Varadhan, H. Yau, \emph{Hydrodynamical  limit for a Hamiltonian system with weak noise}, Commun. Math. Phys. {\bf155}
  (1993), 523-560. 




\bibitem{Va} S.R.S. Varadhan, {\em Nonlinear diffusion limit for a
    system with nearest neighbor interactions-II},  
Asymptotic problems in probability theory: stochastic models and
diffusions on fractals (Sanda/Kyoto, 1990), 75--128, Pitman Res. Notes
Math. Ser., 283, Longman Sci. Tech., Harlow, 1993.  



\end{thebibliography}
\end{document}